\documentclass[aps,prb,twocolumn]{revtex4-2}

\usepackage[]{amsmath}
\usepackage[]{amssymb}
\usepackage{commath}
\usepackage[T1]{fontenc}
\usepackage[utf8]{inputenc}
\usepackage{natbib}
\usepackage[pdftex]{hyperref}
\usepackage[pdftex]{graphicx}
\usepackage{soul}
\usepackage[american]{babel}
\usepackage[caption=false]{subfig}
\DeclareMathOperator{\asinh}{asinh}
\DeclareMathOperator{\atan}{atan}
\DeclareMathOperator{\atanh}{atanh}

\newcommand{\fet}[1]{\mathbf{#1}}

\newcommand{\vd}[2]{\frac{\delta #1}{\delta #2}}
\newcommand{\dd}[2]{\frac{d #1}{d #2}}
\newcommand{\full}{_{-\infty}^{\infty}}

\newcommand{\fv}[1]{\left\langle #1 \right\rangle}
\newcommand{\ex}{\text{ex}}
\newcommand{\eq}{\text{eq}}
\renewcommand{\Re}{\operatorname{Re}}

\begin{document}
\title{Analytical computation of the demagnetizing energy of thin film domain walls}
\author{Audun Skaugen}
\email{audun.skaugen@tuni.fi}
\author{Peyton Murray}
\author{Lasse Laurson}
\affiliation{Helsinki Institute of Physics and Computational Physics Laboratory, Tampere University, P.O. Box 692, FI-33014 Tampere, Finland}

\begin{abstract}
Due to its non-local nature, calculating the demagnetizing field remains the biggest challenge in understanding domain structures in ferromagnetic materials.
Analytical descriptions of demagnetizing effects typically approximate domain walls as uniformly magnetized ellipsoids, neglecting both the smooth rotation of magnetization from one domain to
the other and the interaction between the two domains. Here, instead of the demagnetizing field, we compute analytically the
demagnetizing energy of a straight domain wall described by the classical $\tanh$ magnetization profile in a thin film with perpendicular magnetic anisotropy.
We then use our expression for the demagnetizing energy to derive an improved version of the 1D model of field-driven domain wall motion,
resulting in accurate expressions for important properties of the domain wall such as the domain wall width and the Walker breakdown field.
We verify the accuracy of our analytical results by micromagnetic simulations.
\end{abstract}
\maketitle

\section{Introduction}
Domain walls (DWs) in low-dimensional ferromagnetic systems such as nanowires
and thin films are an active field of study, with promising applications in
spintronics such as memory \cite{Parkin190} and logic \cite{Allwood1688} devices. These
applications typically rely on magnetic fields \cite{Schryer1974,beach2005dynamics,Baltz2007} or spin-polarized electric
currents \cite{Thiaville_2005,Moore2008} to drive DW motion. Hence, accurate analytical and numerical
descriptions of field and current-driven DW dynamics are essential for future device applications.

The basic description of such magnetic systems start with the Landau-Lifshitz-Gilbert
(LLG) equation for the time-evolution of the magnetization vector
${\fet m} = {\fet M}/M_s$, with $M_s$ the saturation magnetization, which in
the case of field-driven magnetization dynamics reads \cite{Gilbert2004}
\begin{equation}
  \pd{\fet m}{t} + \alpha \fet m \times \pd{\fet m}{t} = \gamma\fet m
\times \fet B_{\text{eff}},
  \label{eq:LLG}
\end{equation}
where $\alpha$ is the phenomenological Gilbert damping constant, and
$\gamma$ is the gyromagnetic ratio. The effective field
$\fet B_{\text{eff}}$ in Eq.~(\ref{eq:LLG}) can be
formulated in terms of the total energy $E$ of the system,
\begin{equation}
  \fet B_{\text{eff}} = - \frac{1}{M_s}\vd{E}{\fet m}.
  \label{eq:Beff}
\end{equation}
The energy contains contributions from the exchange, anisotropy, and Zeeman energies, as well as the demagnetizing energy
due to the long-range interaction between magnetization vectors. Numerical
solutions of Eq.~(\ref{eq:LLG}) using a given space discretization are
referred to as micromagnetic simulations, and form an important part of
studies of DWs and their dynamics.

From an analytical perspective, a class of widely used reduced models of
DW dynamics is given by the so-called 1D models, describing the DW in terms
of a smoothly varying one-dimensional magnetization profile parameterized
by the DW position, width, and an angular variable describing the
orientation of the magnetization inside the DW. Dynamical equations for these variables
are derived from the LLG equation
\cite{malozemoff1979magnetic,porter2004velocity,thiaville2006domain,mougin2007domain,slastikov2019walker}.
As a general feature,
such models (as well as the corresponding micromagnetic simulations) exhibit
a regime of steady DW dynamics for small applied fields $B_a$
with the DW velocity increasing with the field.
For fields stronger than a specific driving field magnitude $B_W$, the internal magnetization of the DW starts precessing, resulting in
an abrupt drop in the DW propagation velocity. This instability is related
to the breakdown of the solution found by Schryer and Walker describing the
steady field-driven propagation of an infinitely extended planar DW \cite{Schryer1974},
and is referred to as Walker breakdown.

In ferromagnetic systems with reduced dimensions compared to the DW
width such as nanowires and (ultra) thin films, demagnetizing effects due
to the spatial confinement of the DW become important. The demagnetizing field $\fet B_d = \mu_0\fet H_d$, arising from the demagnetizing part
of the energy in the expression (\ref{eq:Beff}), contains nonlocal contributions from magnetic volume charges $\nabla\cdot \fet m$
and surface charges $\fet m \cdot \fet n$ at the boundary of the system, and gives rise to effects such as shape anisotropy, which penalizes any magnetization normal to the boundary,
and the restoring force, which pulls the DW towards the center of the sample to keep the net magnetization neutral.
In general, the demagnetizing field poses the biggest challenge to understanding domain structures in magnetic systems due to
its long-range nature. A direct computation of the $\fet H_d$ at any given point is often intractable except in very simple cases. One such
case is that of a uniformly magnetized ellipsoid, where the demagnetizing field inside the sample can be given as \cite{hubert2008magnetic}
\begin{equation}
  \fet H_d = -M_s\left( N_x m_x \fet e_x + N_y m_y \fet e_y + N_z m_z \fet e_z \right),
  \label{eq:demagfield}
\end{equation}
where the constants $N_i$, $i=x,y,z$, known as the demagnetizing factors, depend
on the axes of the ellipsoid in question, and must satisfy $N_x + N_y + N_z = 1$.
The simplicity of the demagnetizing factors has motivated approximations
where the demagnetizing field is assumed to follow the form (\ref{eq:demagfield})
even when it is strictly speaking not applicable. For example, in order to study the effects
of the demagnetizing field on DW motion in thin films, Mougin {\it et.\ al.}~\cite{mougin2007domain} modelled
the DW as a uniformly magnetized ellipsoid with axes $(w, D, \delta)$, which (with $\delta \ll D < w$) results in
demagnetizing factors $N_i^E$ given by
\begin{equation}
  N_x^E \approx \frac{\delta}{\delta + w}, \quad N_y^E \approx \frac{\delta}{\delta + D}.
  \label{eq:demag_ellipsoid}
\end{equation}
While this allows a simple description of demagnetizing fields, it is a rather coarse approximation because it ignores both the rapid variation
of magnetization inside the DW as well as the interaction between the two domains and the DW. This directly affects the accuracy of the resulting
properties of the DW, such as the Walker breakdown field $B_W$ and the DW width.

\begin{figure}[tp]
  \subfloat{\includegraphics[width=0.5\textwidth]{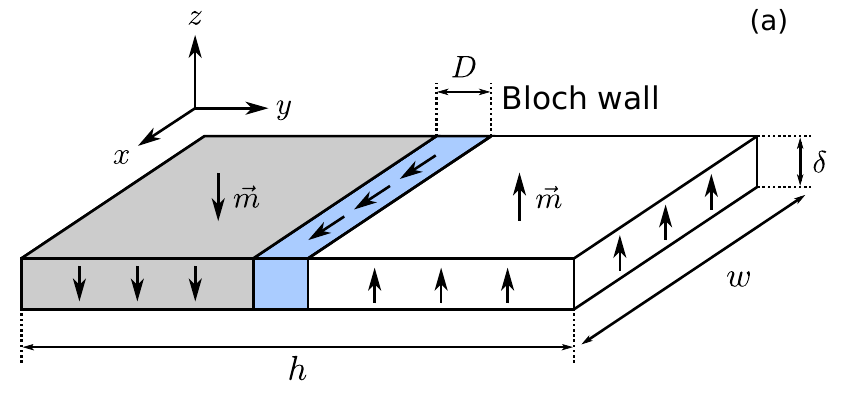} \label{fig:bwa}} \\
  \subfloat{\includegraphics[width=0.5\textwidth]{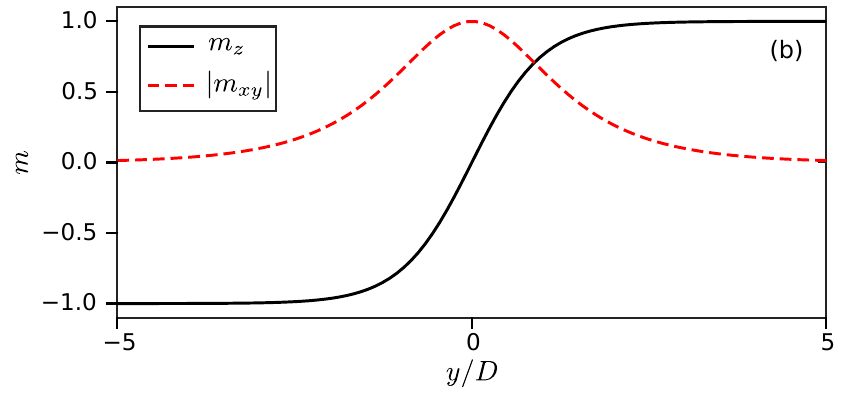} \label{fig:bwb}} 
  \caption{(a): Schematic illustration of the Bloch wall configuration. As we cross the wall, the magnetization rotates from $-z$, via the $x$ direction, to $+z$. We use the form in Eq.~(\ref{eq:wall}), where this rotation is smooth
  as in (b), and the direction of the in-plane magnetization $\fet m_{xy}$ inside the DW can point in any direction, not just $x$.}
  \label{fig:blochwall}
\end{figure}
In this paper, instead of working with the demagnetizing field itself,
we compute the energy due to the demagnetizing field of a uniformly magnetized,
straight DW in a thin film with perpendicular magnetic anisotropy (PMA). As we shall show, the demagnetizing energy is more
analytically tractable than the field, and still lets us derive dynamical equations
for the DW using a Lagrangian framework. We assume a straight, infinitely long DW with no variation in the direction $\fet e_z$ normal to the film, which is valid if the film has thickness $\delta \ll D$. 
The direction of the in-plane magnetization vector, measured by the angle $\phi$ that $\fet m$ makes with the $x$ axis inside the DW, is taken to be uniform.
$\phi = 0$ corresponds to a Bloch wall configuration
(see Fig.~\ref{fig:bwa}), which is energetically preferred due to the absence of magnetic volume charges $\nabla \cdot \fet m$.
However, applying a magnetic field $B_a$ in the $z$ direction will cause the in-plane magnetization to
rotate into the direction $\fet e_y$ normal to the wall, so that a moving DW is associated with
$\phi \neq 0$ (see section \ref{sec:1Dmod}). We therefore keep the value of $\phi$ general in the following.
Uniform DW solutions of the LLG equation (\ref{eq:LLG}), located at $y = Q$, take the general form \cite{hubert2008magnetic}
\begin{equation}
  \fet m(\fet r) = \tanh\left( \frac{y - Q}{D} \right)\fet e_z + \frac{\fet e_x\cos\phi  + \fet e_y\sin\phi}{\cosh\left( \frac{y-Q}{D} \right)},
  \label{eq:wall}
\end{equation}
(see Fig.~\ref{fig:bwb}), where the DW width $D$ remains to be determined. The derivation of this solution ignores the nonlocal effect of demagnetizing fields, however we do not expect deviations from this form to be important.
We will therefore use this form when computing the demagnetizing energy.

The rest of this paper is structured as follows: In Sec.~\ref{sec:energies} we derive a convenient form for the contributions to the demagnetizing energy due to the in-plane and out of plane parts of the magnetization vector,
respectively. We then study each of these contributions separately in the following sections \ref{sec:inplane} and \ref{sec:outofplane}, before applying the results to determine the DW width $D$ in Sec.~\ref{sec:dwwidth},
and to the motion of DWs in Sec.~\ref{sec:1Dmod}. Our results are verified by comparison with micromagnetic simulations in Sec.~\ref{sec:micromagnetics}, before we conclude in Sec.~\ref{sec:conclusion}.

\section{Magnetostatic energy integrals}
\label{sec:energies}
The demagnetizing energy due to a magnet with magnetization vector $\fet m(\fet r)$ is given by
\begin{equation}
  E_d = -\frac{\mu_0 M_s}{2} \int \fet m \cdot \fet H_d d^3\fet r,
  \label{eq:demag_def}
\end{equation}
where the demagnetizing field $\fet H_d$ is determined by Gauss' law for magnetic fields, $\nabla\cdot \fet B = \mu_0 (\nabla \cdot \fet H_d + M_s\nabla\cdot \fet m) = 0$, as well as
Ampere's law $\nabla \times \fet H_d = \fet J = 0$. The solution of these equations can be given in terms of Green's functions as
\begin{align}
  \fet H_d(\fet r) &= \fet H_d^V(\fet r) + \fet H_d^S(\fet r),
  \label{eq:stray-green-tot} \\
  \fet H_d^V(\fet r) &= \frac{M_s}{4\pi} \nabla \int \frac{\nabla'\cdot \fet m'}{|\fet r - \fet r'|}d^3 \fet r',
  \label{eq:stray-green-vol} \\
  \fet H_d^S(\fet r) &= -\frac{M_s}{4\pi} \nabla \int \frac{\fet m'\cdot d \fet S'}{|\fet r - \fet r'|},
  \label{eq:stray-green-surf}
\end{align}
where $\fet m' = \fet m(\fet r')$, $\nabla'$ denotes differentiation with respect to $\fet r'$, and $d\fet S'$ is the surface normal element at the boundary of the system.
In a thin film of thickness $\delta$ much smaller than the relevant magnetic length scales, we can assume that the magnetization
is constant in the $z$ direction normal to the film surface. Taking the film to be large in the lateral directions, the only relevant boundaries are the two horizontal surfaces of the film at
$z = \pm \frac{\delta}{2}$. Inserting the Green's function integrals for $\fet H_d$ into the demagnetizing energy, integrating by parts, and using these assumptions, we can show that the energy takes the form
\begin{align}
  E_d &= E_d^V + E_d^S, \\
  E_d^V &= \frac{\mu_0 M_s^2}{8\pi} \iint (\nabla\cdot \fet m)(\nabla'\cdot\fet m') g_\delta (|\fet r - \fet r'|)d^2 \fet r d^2\fet r', \label{eq:EdV} \\
  E_d^S &= \frac{\mu_0 M_s^2}{8\pi} \iint m_z m'_z f_\delta(|\fet r - \fet r'|)d^2 \fet r d^2\fet r', \label{eq:EdS}
\end{align}
where the integration now extends only over the two-dimensional area of the film. The in-plane interaction kernel $g_\delta$ is given by integrating out the $z$ direction,
\begin{align}
  g_\delta(r) &= \int_{-\frac{\delta}{2}}^{\frac{\delta}{2}}\int_{-\frac{\delta}{2}}^{\frac{\delta}{2}} \frac{dz dz'}{\sqrt{r^2 + (z-z')^2}} \notag \\
  &= 2\delta \asinh \frac{\delta}{r} + 2r - 2 \sqrt{r^2 + \delta^2}.
\end{align}
The out of plane interaction kernel, meanwhile, comes from a surface integral over the thin film boundary, so we must instead evaluate the two coordinate vectors at the upper and lower boundaries,
\begin{align}
  f_\delta(r) &= \left[ \frac{1}{\sqrt{r^2 + (z-z')^2}} \right]_{z,z'= - \frac{\delta}{2}}^{\frac{\delta}{2}}
  = \frac 2 r - \frac{2}{\sqrt{r^2 + \delta^2}}.
\end{align}

We now consider the in-plane $E_d^V$ and the out of plane $E_d^S$ contributions to the demagnetizing energy separately.

\section{In-plane energy and the effective demagnetizing constant}
\label{sec:inplane}
The in-plane demagnetizing energy $E_d^V$ in Eq.~(\ref{eq:EdV}) requires the divergence of the magnetization in Eq.~(\ref{eq:wall}), which is given by
\begin{equation}
  \nabla\cdot \fet m = \pd{m_y}{y} = -\sin \phi \frac{\sinh \frac{y-Q}{D}}{D\cosh^2 \frac{y-Q}{D}}.
\end{equation}
Inserting into Eq.~(\ref{eq:EdV}) and scaling the coordinates by $\frac 1 D$, the interaction kernel $g_\delta$ will transform as $g_\delta(r) = D g_{\frac{\delta}{D}}\left( \frac r D \right)$. Also defining the
small aspect ratio $\sigma = \frac \delta D$, we find
\begin{widetext}
  \begin{align}
	E_d^V = \mu_0M_s^2\sin^2\phi\frac{D^3}{8\pi} \int_{-\frac{w}{2D}}^{\frac{w}{2D}}\int _{-\frac{w}{2D}}^{\frac{w}{2D}}\int_{-\frac{h}{2D}}^{\frac{h}{2D}}\int_{-\frac{h}{2D}}^{\frac{h}{2D}} \frac{\sinh (y-q) \sinh (y'-q)}{\cosh^2 (y-q) \cosh^2 (y'-q)}
	g_{\sigma}(\sqrt{(x-x')^2+(y-y')^2})d x d x' d y d y',
  \end{align}
\end{widetext}
where $w,h$ are the linear sizes of the film in the $x,y$ directions, respectively, and $q = \frac Q D$. This integrand decays exponentially with $y$ and $y'$, so we can take $h\to \infty$ and $q \to 0$ without changing the result
significantly as long as the DW is located near the center of the film.
On the other hand, by symmetry we expect the energy to increase linearly with the width $w$ of the system, so we absorb this and some other constants into the definition by setting $\mathcal E_V = \frac{E_d^V}{\mu_0 M_s^2 w\sin^2\phi}$
and working with the reduced energy $\mathcal E_V$ instead. We now substitute into relative coordinates given by
\begin{align}
  u = x - x', &\quad U = \frac 1 2 (x + x'), \notag \\
  v = y - y', &\quad V = \frac 1 2 (y + y'), \label{eq:relcoord}
\end{align}
which transforms the integration limits to $-\frac{w}{D}..\frac w D$ for the $u$ integral and $- \frac{w-D|u|}{2D}.. \frac{w-D|u|}{2D}$ for the $U$ integral. Since the integrand is independent of $U$, the
integration over this variable amounts to a factor $\frac w D - |u|$.

The hyperbolic functions are most easily transformed to these coordinates by writing them out using their exponential definititions. The resulting transformed integral is given by
\begin{align}
  \mathcal E_V = &\frac{D^2}{4\pi}\int_{-\infty}^\infty \int_{-\infty}^\infty\frac{\cosh 2V - \cosh v}{(\cosh 2V + \cosh v)^2}G_{\sigma}(v)d Vd v, \\
  G_\sigma(v) &= \int_{-\frac{w}{D}}^{\frac w D}\left( 1 - \frac{D}{w}|u| \right)g_{\sigma}(\sqrt{u^2+v^2})d u.
  \label{eq:Gs1}
\end{align}
The integral over $V$ can be done by substituting $t = e^{2V}$ and performing a partial fraction decomposition. Using that the integrand is even in $v$ to restrict the limits to $0\ldots\infty$ and
integrating by parts in $v$, we obtain
\begin{align}
  \mathcal E_V = -\frac{D^2}{2\pi}\int_{0}^\infty \frac{v\cosh v - \sinh v}{\sinh^2 v}\pd{G_{\sigma}(v)}{v}d v.
\end{align}
After differentiation with respect to $v$ under the integral sign and taking $w \to \infty$, the integral over $u$ in the $G_\sigma$ function can be computed, giving
\begin{equation}
  \pd{G_\sigma}{v} = 4\sigma\atan \frac{v}{\sigma} - 2\sigma\pi - 4v\ln \frac{v}{\sqrt{v^2 + \sigma^2}}.
  \label{eq:Gs2}
\end{equation}
This is further simplified by another differentiation with respect to $v$,
\begin{equation}
  \pd[2]{G_\sigma}{v} = -4\ln \frac{v}{\sqrt{v^2 + \sigma^2}} = 2\ln \left( 1 + \frac{\sigma^2}{v^2} \right),
  \label{eq:Gs3}
\end{equation}
so after another integration by parts the energy is simplified to
\begin{equation}
  \mathcal E_V = \frac{D^2}{\pi}\left[ \sigma\pi - \int_{0}^\infty \frac{v}{\sinh v}\ln \left( 1 + \frac{\sigma^2}{v^2} \right)d v\right].
  \label{eq:EVexpand}
\end{equation}
To make further progress we will need to expand in the small parameter $\sigma$ and integrate term by term. However, a naive expansion of Eq.~(\ref{eq:EVexpand}) leads to integrals which diverge at the origin. This is because the
interchange of summation and integration is only valid if the integrand is an analytic function of $v$ on the entire contour of integration,
but the integrand has a branch cut when $v$ goes from $-i\sigma$ to $i\sigma$, which includes $v=0$. In order to avoid this branch cut, we extend the integration limits back to $-\infty\ldots\infty$ and
decompose the logarithm as
\begin{align}
  \mathcal E_V &= \delta D - \frac{D^2}{2\pi}\int_{-\infty}^\infty \frac{v}{\sinh v}\left[ \ln\left( 1 + i\frac\sigma v \right) + \ln\left( 1 - i\frac \sigma v \right) \right]d v \notag\\
  &= \delta D + \frac{D^2}{\pi}\Re I_\sigma,
\end{align}
\begin{figure}[tp]
  \includegraphics[width=0.5\textwidth]{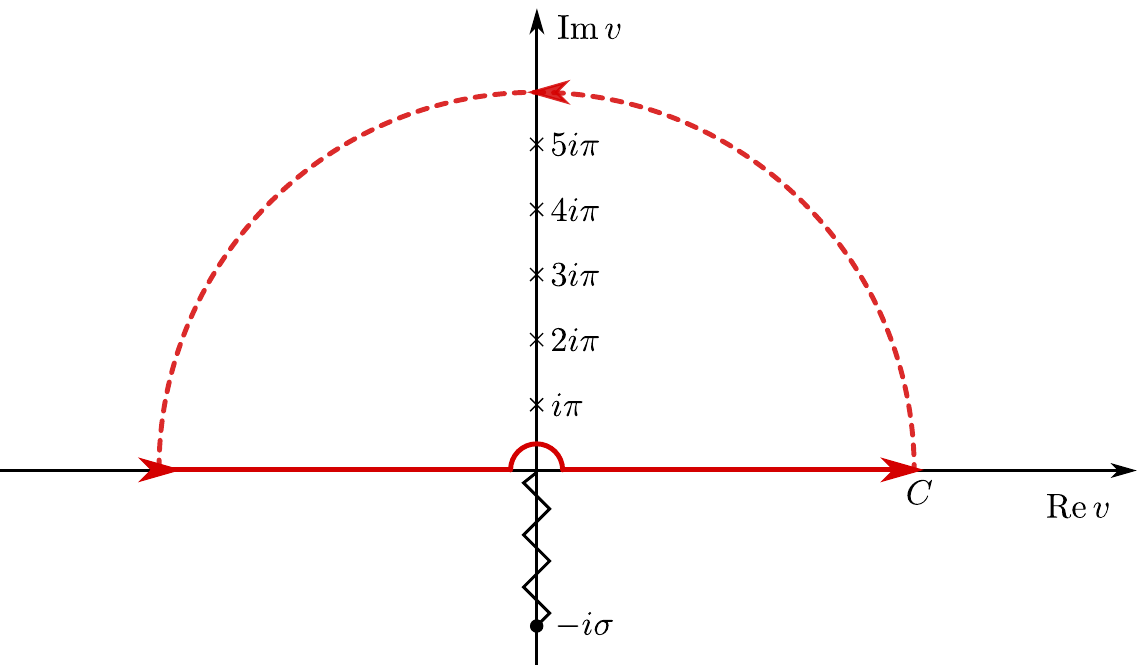}
  \caption{Analytical structure of the integrand of $I_\sigma$. The function $\ln\left( 1 + \frac{\sigma^2}{v^2} \right)$ has branch cuts going from the origin to $\pm i\sigma$. One of these is removed by factoring the argument to
	the logarithm and looking at the $\ln\left( 1 + i\frac \sigma v \right)$ part, and is not shown in the figure. The other (zigzag line) is avoided by deforming the integration contour into the positive imaginary half-plane (solid red line).
	The $\frac{1}{\sinh v}$ function has poles at $v = i\pi k$ (crosses). After series expansion and regularization, the integrals are evaluated by extending the contour around the positive imaginary half-plane (dashed red line)
and using the residue theorem.}
  \label{fig:contour}
\end{figure}
where $I_\sigma$ is the integral keeping only the first term inside the square brackets. This isolates the branch cut to the negative imaginary half-plane, so we can deform the integration contour to the contour $C$ going
from $-\infty$ to $-r$ with $r>0$ an arbitrarily small number, then around a semicircle of radius $r$ into the positive imaginary half-plane to avoid the origin, then from $r$ to $\infty$ (see Fig.~\ref{fig:contour}). Expanding the logarithm in $\sigma$, we find
\begin{align}
  I_\sigma = \sum_{n=1}^\infty \frac{(-i\sigma)^n}{n} I_n, \quad
  I_n = \int_C \frac{v}{\sinh v}v^{-n}d v. \label{eq:EdV_deformed}
\end{align}
These integrals can be solved by multiplying the integrand with $e^{i\epsilon v/\pi}$ for some $\epsilon>0$ to ensure convergence in the positive imaginary half-plane, then extending the integration contour with a counterclockwise semicircle
of radius $R$, which gives a vanishing contribution when $R \to \infty$. Summing over the residues at $v = i\pi k, k\in \mathbb N$ and then taking $\epsilon \to 0$, we find the values
\begin{align}
  I_1 &= 2\pi i\lim_{\epsilon\to 0} \sum_{k=1}^\infty (-e^{-\epsilon})^k = i\pi, \\
  I_2 &= 2\lim_{\epsilon\to 0}\sum_{k=1}^\infty \frac 1 k \left( -e^{-\epsilon} \right)^k = -2\ln 2, \\
  I_n &= 2 (i\pi)^{2-n}\sum_{k=1}^\infty \frac{(-1)^k}{k^{n-1}} \notag \\
  &= -2(i\pi)^{2-n}\left( 1 - 2^{2-n} \right)\zeta(n-1),
\end{align}
where the last line, valid for $n>2$, uses a known relation between the Dirichlet eta function $\eta(s) = \sum_{k=1}^\infty \frac{(-1)^{s+1}}{k^s}$, and the Riemann zeta function \cite{etafunction}. Inserting back into $\mathcal E_V$, the first-order term will
cancel with the $\delta D$ term, giving an in-plane reduced energy of
\begin{align}
  \mathcal E_V &= \frac{\delta^2}{\pi}\left[ \ln 2 + \sum_{n=3}^\infty \frac{2(-\sigma)^{n-2}}{n \pi^{n-2}}\left( 1-2^{2-n} \right)\zeta(n-1) \right] \notag \\
&= \left( \frac{\delta^2}{\pi} \ln 2 - \frac{\delta^3}{18D} + \frac{3\delta^4}{8\pi^3D^2}\zeta(3) \right) + \delta^2 \mathcal O\left( \sigma^3 \right),
\label{eq:EdVFinal}
\end{align}
recalling that the full demagnetizing energy is related to this quantity by $E_d^V = \mu_0 M_s^2 w \sin^2\phi \mathcal E_V$.

This energy can be interpreted in terms of an effective demagnetizing constant $N_y$ inside the DW. Such a demagnetizing constant would mean that the demagnetizing field is given by
\begin{equation}
  H^d_y = -M_s N_y m_y = -M_s N_y \frac{\sin\phi}{\cosh\left( \frac{y-q}{D} \right)}.
\end{equation}
Inserting into Eq.~(\ref{eq:demag_def}), this leads to a demagnetizing energy given by
\begin{align}
  \frac{E_d}{w} &= \frac 1 2 \mu_0 M_s^2N_y \delta \sin^2\phi \int_{-\infty}^\infty \frac{dy}{\cosh^2 \left( \frac{y-q}{D} \right)} \notag\\
  &= \mu_0 M_s^2 N_y \delta D \sin^2\phi. \label{eq:EdV_demag}
\end{align}
Comparing with the energy we computed in Eq.~(\ref{eq:EdVFinal}), we see that $N_y$ must be chosen as
\begin{equation}
  N_y = \frac{\delta}{\pi D}\ln 2 - \frac{\delta^2}{18 D^2} + \mathcal O\left( \sigma^3 \right).
\label{eq:Ny}
\end{equation}

This expression should be compared with the demagnetizing constant obtained by taking the DW as a uniformly magnetized ellipsoid, given in Eq.~(\ref{eq:demag_ellipsoid}), which can be expanded in $\delta/D$ to give
\begin{equation}
  N_y^E = \frac \delta D - \frac{\delta^2}{D^2} + \frac{\delta^3}{D^3} - \ldots.
\end{equation}
While this has the same qualitative behavior as our expression (\ref{eq:Ny}), quantitatively it is quite different. Our lowest order term is smaller by a factor $\frac{\ln 2}{\pi} \approx 0.22$, which has a direct effect on
the motion of DWs (see Sec.~\ref{sec:1Dmod}). Indeed, in Ref.~\cite{Baltz2007}, the elliptic approximation was used to estimate the Walker field $B_W$ from experimentally measurable quantities,
giving $B_W \approx 12\,\text{mT}$ for the
$0.5\,\text{nm}$ thin film, while micromagnetic simulations of the same system instead gave $B_W \approx 2.7\,\text{mT}$ \cite{herranen2019barkhausen}, which is reproduced by our analytical computation
(see also Sec.~\ref{sec:micromagnetics}). Other authors use $\pi D$ in place of $D$ in the expression for $N_y^E$ \cite{BOULLE2011159}. This gets closer to our result, but will still give a different second-order correction.

\section{Out of plane energy and the restoring force}
\label{sec:outofplane}
Inserting $m_z = \tanh(\frac{y-Q}{D})$ into Eq.~\ref{eq:EdS} and scaling the coordinates by $1/D$, the interaction kernel transforms as $f_\delta(r) = \frac 1 D f_\sigma\left(\frac r D\right)$, giving
\begin{widetext}
  \begin{equation}
	E_d^S = \mu_0 M_s^2\frac{D^3}{8\pi}\int_{-\frac{w}{2D}}^{\frac{w}{2D}}\int_{-\frac{w}{2D}}^{\frac{w}{2D}}\int_{-\frac{h}{2D}}^{\frac{h}{2D}}\int_{-\frac{h}{2D}}^{\frac{h}{2D}}
	\tanh(y-q)\tanh(y'-q)f_{\sigma}\left( \sqrt{(x-x')^2 + (y-y')^2} \right)dx dx' dy dy',
  \end{equation}
\end{widetext}
where again $q = \frac Q D$. By contrast to the in-plane energy, this energy is not only due to the DW, but also contains significant contributions from the dipole charges of the domains themselves. We will therefore
have to keep the system size $h$ and the position $q$ general, expecting in particular a quadratic dependence on the position $q$ for a linear restoring force.
Changing variables to the relative coordinates of Eqs.~(\ref{eq:relcoord}), taking $w \to \infty$
and integrating over the $U, u$ variables, the reduced energy $\mathcal E_S = \frac{E_d^S}{\mu_0 M_s^2 w}$ takes the form
\begin{align}
  &\mathcal E_S = \frac{D^2}{8\pi}\int_{0}^{\frac{h}{D}}\!\!\!\int_{-\frac{h-Dv}{2D}}^{\frac{h-Dv}{2D}}\frac{\cosh 2(V-q)-\cosh v}{\cosh 2(V-q)+\cosh v} F_\sigma(v)dV dv,\notag \\
  &F_\sigma(v) = \int_{-\infty}^\infty f_\sigma(\sqrt{u^2+v^2})\dif u = 4\ln \left( 1 + \frac{\sigma^2}{v^2} \right), \label{eq:Fs1}
\end{align}
where we also used that the integrand is even in $v$ to keep $v$ positive while integrating, simplifying sign issues.
The integral over $V$ can now be done using similar techniques as for the in-plane energy. However, the more general limits of integration lead to a more complicated expression. Defining the small aspect ratio $\nu = \frac D h$, we find
\begin{align}
  &\mathcal E_S = \frac{D^2}{8\pi}\int_0^{\nu^{-1}}\left[ \coth v M(v) + \nu^{-1} - v \right]\! F_\sigma(v)\,dv, \label{eq:EsVint}\\
  &M(v; \nu, q) = \ln \left[ \frac{\cosh (\nu^{-1} - 2v) + \cosh(2q)}{\cosh(\nu^{-1}) + \cosh(2q)} \right].
\end{align}
If the DW is close to the center of the sample, we will have $q \ll \nu^{-1}$, so the denominator of the logarithm can be simplified to $\cosh (\nu^{-1})$.
We can then expand in $q$ to give $M(v) = M_0(v) + M_2(v) q^2$ with
\begin{align}
  M_0(v) &= \ln\left[ \frac{\cosh(\nu^{-1}-2v)+1}{\cosh(\nu^{-1})} \right],  \label{eq:M0}\\
  M_2(v) &= \frac{2}{\cosh(\nu^{-1}-2v)+1}.
\end{align}
In this way, the energy is decomposed as $\mathcal E_S = \mathcal E_M + \mathcal E_q + \mathcal E_a$, with
\begin{align}
  \mathcal E_M &= \frac{D^2}{2\pi}\int_0^{\nu^{-1}}\coth v M_0(v)\ln\left( 1+ \frac{\sigma^2}{v^2} \right)dv,\\
  \mathcal E_q &= \frac{Q^2}{2\pi}\int_0^{\nu^{-1}}\coth v M_2(v)\ln\left( 1+\frac{\sigma^2}{v^2} \right)dv, \\
  \mathcal E_a &= \frac{D^2}{2\pi}\int_0^{\nu^{-1}}(\nu^{-1}-v)\ln\left( 1 + \frac{\sigma^2}{v^2} \right).
\end{align}
We first consider the position-dependent $\mathcal E_q$, which will determine the strength of the restoring force pulling the DW towards the center of the sample.
We can note that $M_2$ is exponentially small unless $v$ is close to $\frac{\nu^{-1}}{2}$, so we can extend the limits of integration to infinity. Changing variables to $t = v-\frac 1 2\nu^{-1}$, we find
\begin{align}
  \mathcal E_q &= \frac{Q^2}{2\pi}\int_{-\infty}^\infty \frac{2}{\cosh(2t)+1}\ln \left( 1 + \frac{4\sigma^2\nu^2}{(1 + 2\nu t)^2} \right)dt \notag \\
  &= \frac{Q^2}{\pi}\int\full\frac{1}{\cosh(2t)+1}\left( 4\nu^2\sigma^2 + \mathcal O(\nu^4) \right) dt \notag \\
  &= \frac{4\delta^2 Q^2}{\pi h^2} + \delta^2\mathcal O\left( \nu^4 \right),
  \label{eq:restoring_quadratic}
\end{align}
where we expanded the logarithm in powers of $\nu$ and kept the lowest-order term. We will want to compute the other energy terms to the same order in $\nu$, so that we can compare energies accurately.
Of these, $\mathcal E_a$ can be done simply using integration by parts, giving
\begin{align}
  \mathcal E_a &= 
  \frac{\delta h}{2} + \frac{\delta^2}{2\pi}\ln\left( \frac{\delta}{h} \right) - \frac{3\delta^2}{4\pi} - \frac{\delta^4}{24\pi h^2} + D^2\mathcal O(\nu^4),
\end{align}
where we expanded the result to order $\nu^2$.

\begin{figure}[tp]
  \includegraphics[width=0.5\textwidth]{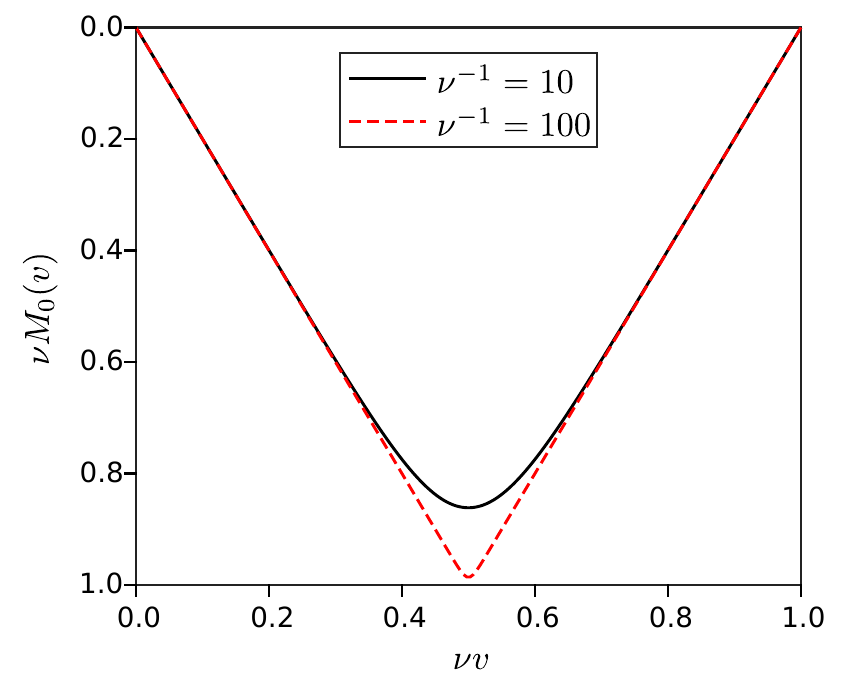}
  \caption{Illustration of the $M_0(v)$ function defined in Eq.~(\ref{eq:M0}). The function is symmetric about the $v = \frac{\nu^{-1}}{2}$ point, and behaves linearly far away from this point. As $v$ approaches the midpoint, the function deviates from
  linear behavior at a rate dictated by the size of $\nu$.}
  \label{fig:monster}
\end{figure}
To compute $\mathcal E_M$, it helps to understand the $M_0(v)$ function. It is plotted in Fig.~\ref{fig:monster}, where we see that it can be approximated well
by two different linear functions of $v$ for the first and second half of the interval, respectively. Indeed, writing out the hyperbolic cosine and approximating $2\cosh\nu^{-1} \approx e^{\nu^{-1}}$, we find
\begin{align}
  M_0(v) &= \ln \left( e^{\nu^{-1}-2v} + e^{2v-\nu^{-1}} + 2 \right) - \nu^{-1} \\
  &= \begin{cases}
	-2v + 2\ln \left( e^{2v-\nu^{-1}} + 1 \right) & v < \frac{\nu^{-1}}{2} \\
	2(v-\nu^{-1}) + 2\ln\left( e^{\nu^{-1}-2v} + 1 \right) & v > \frac{\nu^{-1}}{2}
  \end{cases}, \notag
\end{align}
where deviations from the linear behavior, described by the logarithms, are significant only when $v$ is close to $\frac{\nu^{-1}}{2}$.
We therefore decompose this energy further as $\mathcal E_M = \mathcal E_L + \mathcal E_R + \mathcal E_d$, where $\mathcal E_L$ and $\mathcal E_R$ capture the linear behavior at the left and right half, respectively,
and $\mathcal E_d$ captures the deviations from linear behavior on both sides. Ignoring the $\coth v$ function in $\mathcal E_d$ and $\mathcal E_R$, this means that
\begin{align}
  \mathcal E_L &= -\frac{D^2}{\pi}\int_0^{\frac{\nu^{-1}}{2}}v\coth v \ln\left( 1+\frac{\sigma^2}{v^2} \right)dv, \\
  \mathcal E_R &= \frac{D^2}{\pi}\int_{\frac{\nu^{-1}}{2}}^{\nu^{-1}}(\nu^{-1}-v) \ln\left( 1+\frac{\sigma^2}{v^2} \right)dv, \\
  \mathcal E_d &= \frac{D^2}{\pi}\int_0^{\nu^{-1}}\ln\left( 1 + e^{-|\nu^{-1}-2v|} \right)\ln\left( 1+\frac{\sigma^2}{v^2} \right)dv.
\end{align}
For $\mathcal E_d$ we use the same substitution and expansion as we did for $\mathcal E_q$ to find
\begin{align}
  \mathcal E_d &= \frac{D^2}{\pi}\int_{-\infty}^\infty \ln (1 + e^{-2|t|})\left( 4\sigma^2\nu^2 + \mathcal O(\nu^4) \right)dt \notag \\
  &= \frac{\pi \delta^2D^2}{3h^2} + D^2\mathcal O\left(\nu^4\right).
\end{align}
$\mathcal E_R$ can be treated similarly to $\mathcal E_a$, giving
\begin{align}
  \mathcal E_0^R &= \frac{D^2}{\pi}\int_{\frac{\nu^{-1}}{2}}^{\nu^{-1}} (v-\nu^{-1})\ln \left( 1 + \frac{\sigma^2}{v^2} \right)dv \notag \\
  &= \frac{\delta^2}{\pi}\ln 2 - \frac{\delta^2}{\pi} + \frac{5\delta^4}{12\pi h^2} + D^2\mathcal O\left(\nu^4\right).
\end{align}
Finally, in $\mathcal E_L$ we can not ignore the $\coth v$ function. Instead we expand in $\sigma$ using the same techniques as we used for the in-plane energy, namely
\begin{align}
  \mathcal E_0^L &= -\frac{D^2}{2\pi}\int_{-\frac{\nu^{-1}}{2}}^{\frac{\nu^{-1}}{2}}v\coth v \ln \left( 1 + \frac{\sigma^2}{v^2} \right)dv \notag \\
  &= - \frac{D^2}{\pi}\Re I_\sigma,
\end{align}
where $I_\sigma$ replaces the logarithm with $\ln \left( 1 + i \frac{\sigma}{v} \right)$. Deforming the contour to avoid the origin and expanding in $\sigma$, we obtain
\begin{align}
  I_\sigma &= -\sum_{n=1}^\infty \frac{(-i\sigma)^n}{n}I_n, \quad I_n = \int_C v^{1-n}\coth vdv, \label{eq:EdS_deformed}
\end{align}
where the contour $C$ is as in the previous section except that the endpoints are at $\pm \nu^{-1}$ instead of $\infty$.
This still gives significant contributions from large $v$ when $n$ is small, so we can not use the standard residue method here. Instead we treat the integral explicitly by dividing the contour into two parts
along the real line and one small semicircle where we can substitute $v = re^{i\theta}$ and expand in $r$. This results in the values
\begin{align}
  I_1 = -i\pi, \quad I_2 = 2(\lambda - 1 - \ln 2\nu),
\end{align}
where $\lambda$ is a numerical integration constant given by
\begin{align}
  \lambda &= \int_0^1 \left( \frac{\coth v}{v}-v^{-2} \right)dv + \int_1^\infty (\coth v - 1)v^{-1}dv \notag\\
  &\approx 0.4325.
\end{align}
For $n > 2$, we can enlarge the integration contour to infinity and use the standard semicircle contour to find \[
I_n = -2(-i)^n\pi^{2-n}\zeta(n-1) + \kappa_n, \]
where $\kappa_n$ is the error due to enlarging the contour. This vanishes for $n$ odd due to the integrals on the real line cancelling each other, but for even values it can be  approximated as
\begin{equation}
  \kappa_n = -2\int_{\frac{\nu^{-1}}{2}}^\infty v^{1-n}dv = \frac{2^{n-1}}{2-n}\nu^{n-2},
\end{equation}
where we again ignored the $\coth$ function in the integral. To order $\mathcal O(\nu^2)$, only the $n=4$ value is important, with value $\kappa_4 = -4\nu^2$.

Combining everything, the reduced out of plane energy is given to order $\nu^2$ and $\sigma^4$ by
\begin{align}
  \mathcal E_S =& \mathcal E_L + \mathcal E_R + \mathcal E_d + \mathcal E_q + \mathcal E_a \notag \\
  =& \frac{\delta h}{2} - \delta D + \frac{\delta^2}{2\pi}\ln \left(\frac{16\delta D^2}{h^3}\right) - \frac{(3+4\lambda)\delta^2}{4\pi} \notag \\
  &- \frac{5\delta^4}{8\pi h^2} + \frac{\pi\delta^2D^2}{3h^2} + \frac{\delta^3}{9D} - \frac{\zeta(3)\delta^4}{2\pi^3 D^2} + \frac{4\delta^2 Q^2}{\pi h^2}. \label{eq:EdSFinal}
\end{align}
These energy terms can be divided into four different types:
\begin{enumerate}
  \item Terms independent of $D$ and $Q$: These are dominated by the $\frac{\delta h}{2}$ term corresponding to the energy $\frac 1 2 \mu_0 M_s^2 V$ of two isolated domains magnetized in the $z$ direction,
	with combined volume $V = \delta h w$. This of course ignores the excluded volume from the DW itself, which is accounted for by the $-\delta D$ term. Higher-order terms give corrections to this, resulting in an effective
	demagnetizing constant $N_z$ which is slightly below $1$.
  \item The logarithmic term: This can be interpreted as an interaction between the two thin film domains.
	They will attract each other due to the oppositely directed magnetizations, with a force $-\pd{E}{D} \sim -\frac{\delta^2}{D}$.
  \item Terms depending on $D$, but independent of $Q$: These correspond to the demagnetizing energy of the DW itself.
  \item The term proportional to $Q^2$: This represents a harmonic restoring force pulling the DW back to the center at $Q=0$, so that zero net magnetization is preferred.
\end{enumerate}
In refs.~\cite{Urbach1995,Zapperi1998}, the energy due to the restoring force is assumed to take the form
\begin{equation}
  E_Q^N = -\frac 1 2 \mu_0 M_s V \fv{H_d}\fv{m_z} = \frac 1 2 \mu_0 M_s^2 V \mathcal N \fv{m_z}^2,
\end{equation}
where $\fv{-}$ denotes averaging over space, and $\fv{H_d}$ is taken as $-M_s\mathcal N \fv{m_z}$ for some effective demagnetizing constant $\mathcal N$ describing the entire domain structure. Inserting for $m_z$, this gives
\begin{equation}
  E_Q^N = 2\mu_0 M_s^2 \mathcal N\frac{w \delta}{h} Q^2.
\end{equation}
Comparing with our result $E_Q = 4\mu_0 M_s^2 \frac{w\delta^2}{\pi h^2}Q^2$, we see that the effective demagnetizing constant must be chosen as $\mathcal N = \frac{2\delta}{\pi h}$.

\section{Steady-state domain wall width}
\label{sec:dwwidth}
The DW width $D$ is a dynamical variable evolving with time. In equilibrium or steady-state motion, the steady-state DW width is the one that minimizes the energy at a given value of $\phi$.
In addition to the demagnetizing energies we computed above, the Landau-Lifshitz energy includes contributions from the exchange energy and the anisotropy energy.
Using the form of Eq.~(\ref{eq:wall}) for the DW, these energies are readily computed as
\begin{align}
  E_{\ex} &= A_{\text{ex}}\int |\nabla \fet m|^2 d^3 \fet r = 2\delta w \frac{A_{\ex}}{D}, \\
  E_a &= K_u \int (1-m_z^2) d^3 \fet r = 2\delta w D K_u,
\end{align}
where we added a constant energy density to $E_a$ to keep it finite when $h \to \infty$. To the lowest order in $\delta$, the only contributing term of the demagnetizing energy is $\delta D$ from the out of plane
energy, giving a minimizing equation
\begin{align}
  \frac{1}{\delta w}\dpd{E}{D} &= - \frac{2A_\ex}{D^2} + 2K_u - \mu_0 M_s^2 = 0,
\end{align}
with solution $D_0$ given by
\begin{align}
  D_0 &= \sqrt{\frac{A_\ex}{K_u - \frac 1 2 \mu_0 M_s^2}}. \label{eq:D0}
\end{align}
Considering higher orders in $\delta$, it is convenient to define the exchange length $D_\ex = \sqrt{\frac{2A_\ex}{\mu_0 M_s^2}}$. In terms of $D_0$ and $D_\ex$, the minimizing equation is given to order
$\delta^3$ by
\begin{align}
  0 &= \frac{D^2}{2\delta w A_\ex} \dpd{E}{D} \label{eq:dEdD} \\
  &= \frac{D^2}{D_0^2} - 1 + D_\ex^{-2}\left[ \frac \delta \pi D + \frac{\delta^2}{18}(\sin^2\phi - 2) + \frac{2\pi \delta D^3}{3h^2} \right]. \notag
\end{align}
Instead of solving this cubic equation directly, we treat it perturbatively in orders of $\delta$, by expanding the solution $D_\eq$ as 
\begin{equation}
  D_\eq = D_0 + D_1 \delta + D_2 \delta^2 + \mathcal O(\delta^3),
  \label{eq:D_eq}
\end{equation}
and solving for each order in $\delta$ separately.
To order $n=0$, this gives the value in Eq.~(\ref{eq:D0}), while for higher orders we find
\begin{align}
  D_1 &= - \frac{D_0^2}{2\pi D_\ex^2}\left( 1 + \frac{2\pi^2D_0^2}{3h^2} \right), \label{eq:D1} \\
  D_2 &= -\frac{D_1^2}{2D_0} - \frac{D_0}{2 D_\ex^2}\left( \frac{D_1}{\pi} + \frac{\sin^2\phi-2}{18} + \frac{2\pi D_0^2D_1}{h^2} \right), \label{eq:D2}
\end{align}
This can be compared with the result commonly obtained by using demagnetizing constants
(setting $N_x^E=0$ and $N_z^E = 1 - N_y^E$ in this geometry) \cite{mougin2007domain},
\begin{equation}
  D_N = \sqrt{\frac{A_{\ex}}{K_u - \frac 1 2 \mu_0 M_s^2(1 - N_y^E - N_y^E \sin^2\phi)}},
\end{equation}
which depends on the film thickness $\delta$ through $N_y^E = \frac{\delta}{\delta + D_N}$. Expanding to first order in $\delta$, we obtain
\begin{equation}
  D_N = D_0 - \delta\frac{D_0^2}{D_\ex^2}(\sin^2\phi+1) + \mathcal O(\delta^2).
\end{equation}
Here the dependence on the angle $\phi$ is of order $\mathcal O(\delta)$, by contrast with our result which only depends on $\phi$ in the second order term $D_2$.
The difference is that the derivation of $D_N$ ignores the fact that
$N_y$ depends on $D$ when minimizing the energy. As we can see from Eq.~(\ref{eq:EdV_demag}), the in-plane demagnetizing energy due to $N_y$ is proportional to $D N_y$, which should be differentiated with respect to $D$ to find the minimum.
However, the lowest order term of $N_y$ is generally proportional to $\frac \delta D$, so the lowest order term of $\pd{(DN_y)}{D}$ cancels out, leaving only a term of order $\mathcal O(\delta^2)$. Our expression for
$D_\eq$ takes proper account of the dependence of the energy on the DW width, giving the correct dependence on $\phi$ as well as more precise constants.

\section{Domain wall dynamics}
\label{sec:1Dmod}
The demagnetizing energies we computed here can be used to derive an accurate 1D model for the motion of a uniform DW, as originally derived by Slonczewski \cite{malozemoff1979magnetic}.
Following Refs.~\cite{thiaville2006domain,thiaville2004domain}, we employ the Lagrangian formulation of the LLG equation.
The conservative Landau-Lifshitz equation can be posed in a Lagrangian form for the
angle $\theta$ between $\fet m$ and the $z$ axis, and the angle $\phi$ describing the in-plane component. The Lagrangian is given by $L = \frac{M_s}{\gamma}\int \dot \phi \cos \theta d^3\fet r - E$, where the energy
$E$ also includes the Zeeman energy due to a constant applied field $B_a$ in the $z$ direction, given by $-M_s B_a \int m_z d^3 \fet r$. Gilbert dissipation
can be included using a Rayleigh dissipation functional given by $F = \frac{\alpha M_s}{2\gamma}\int |\dot{\fet{m}}|^2d^3 \fet r$. Inserting the ansatz (\ref{eq:wall}) into these functionals and integrating over space, we find the Lagrangian and
dissipation functional governing the three variables $s_i = \set{Q, \phi, D}$ describing the DW. These variables obey the dissipative Euler-Lagrange equation for each variable,
\begin{equation}
  \dd{}{t}\pd{L}{\dot s_i} - \pd{L}{s_i} + \pd{F}{\dot s_i} = 0,
\end{equation}
which results in the equations of motion given by
\begin{align}
  \alpha \frac{\dot Q}{D} + \dot \phi &= -\gamma (B_a + B_R), \quad B_R = \frac{4\mu_0 M_s\delta}{\pi h^2} Q, \label{eq:Q}\\
  \frac{\dot Q}{D} - \alpha \dot \phi &= \frac \gamma 2 \mu_0 M_s N_y \sin(2\phi), \label{eq:phi}\\
  \frac{\dot D}{D} &= -\frac{6\gamma}{\pi^2\alpha M_s w\delta} \dpd{E}{D}. \label{eq:D}
\end{align}
Here $B_R$ is the effective field corresponding to the restoring force, and $N_y$ and $\pd{E}{D}$ are given in
Eqs.~(\ref{eq:Ny}) and (\ref{eq:dEdD}), respectively. Eq.~(\ref{eq:D}) describes how the DW width $D$ relaxes towards the steady-state value $D_\eq$. We can estimate how fast this
relaxation is by linearizing around the steady state. To zeroth order in $\delta$, this gives an exponential approach with relaxation time
\begin{equation}
  \tau_D = \frac{\alpha \pi^2 M_s D_0^2}{24\gamma A_\ex} + \mathcal O(\delta), 
\end{equation}
with higher-order corrections derivable. This timescale is shorter than the timescale $\tau_V = \frac{D}{V_W}$ of fast DW motion (see below) by a factor $\frac{\tau_D}{\tau_V} \propto \alpha N_y \ll 1$.
It is therefore common to ignore the dynamics of $D$ and set $D = D_\eq(\phi)$ at each point in time \cite{thiaville2006domain}.

Walker-like steady-state solutions are found by setting $\dot \phi = 0$ in Eqs.~(\ref{eq:Q}--\ref{eq:phi}). The resulting equations are solvable only if
\begin{equation}
  |B_a + B_R| \le B_W = \frac{\alpha}{2}\mu_0 M_s N_y,
  \label{eq:walkerfield}
\end{equation}
giving an expression for the Walker breakdown field $B_W$. Below this field, the steady-state solution
gives a DW velocity of
\begin{equation}
  \dot Q = -\frac{\gamma D_\eq}{\alpha}(B_a + B_R).
  \label{eq:speed}
\end{equation}
In particular, the velocity at Walker breakdown is given by
\begin{equation}
  V_W = |\dot Q(B_W)| = \frac{\gamma D_\eq}{2}\mu_0 M_s N_y.
  \label{eq:walkerspeed}
\end{equation}
Note that these quantities depend on the geometry through both the steady-state DW width $D_\eq$ (\ref{eq:D_eq}), and the effective demagnetizing constant $N_y$ (\ref{eq:Ny}).

\section{Numerical verification}
\label{sec:micromagnetics}
\begin{figure}[tp]
    \includegraphics[height=7cm]{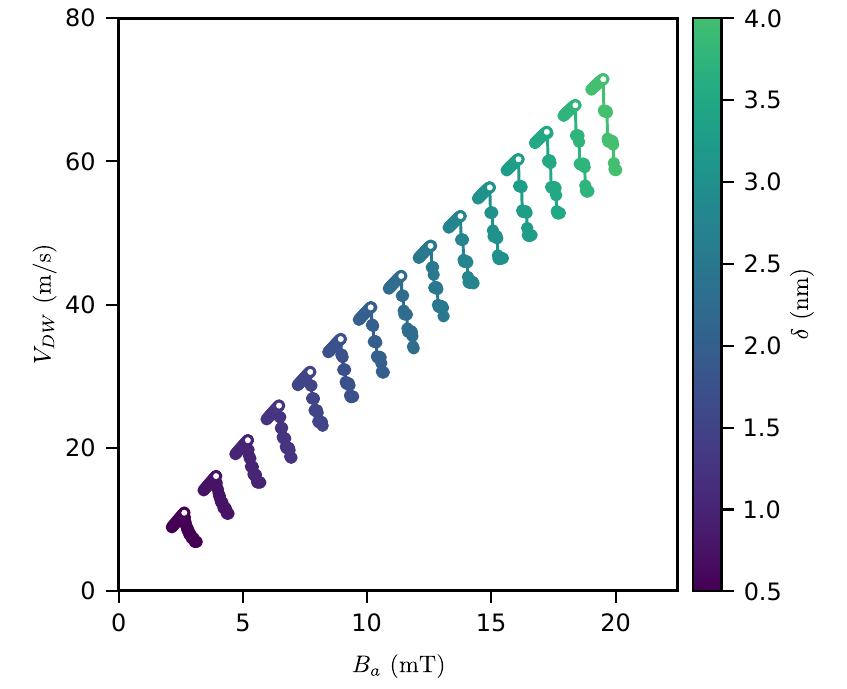}
    \caption{Domain wall velocity as a function of external field for varying film thicknesses, indicated by the color scale.
	  Filled symbols correspond to the average DW velocity at a given applied field $B_a$. The peak velocity at a given thickness (open symbols) gives an estimate for the Walker breakdown field $B_W$
	and velocity $V_W$.}
\label{fig:vdwvsbz}
\end{figure}

To verify our analytic computations, we performed micromagnetic simulations using the MuMax3 software package~\cite{Vansteenkiste2014}. An initial Bloch-type DW configuration was generated by setting the magnetization to point along $+z$ for $0 \le y < h/2$ and $-z$ for $h/2 < y \le h$, with a small region pointing along $+x$ at the boundary between the domains (see Fig.~\ref{fig:bwa}). After relaxing the initial configuration in zero field to an energy minimum, we applied external magnetic field and calculated the subsequent DW motion by numerically integrating the LLG equation (\ref{eq:LLG}). In all simulations we used micromagnetic parameters previously determined from experiments on Pt/Co/Pt films with perpendicular anisotropy~\cite{Baltz2007}, with exchange stiffness $A_{\ex} = 1.4 \times 10^{-11} \, \text{J/m}$, saturation magnetization $M_s = 9.1 \times 10^5 \, \text{A}/\text{m}$, uniaxial anisotropy constant $K = 8.4 \times 10^5 \, \text{J}/\text{m}^3$, and dissipation constant $\alpha = 0.27$.

The variation of the Walker breakdown field $B_W$ with the film thickness $\delta$ was computed on a rectangular grid of $64\times 128\times 1$ cells with in-plane cell size $\Delta x = \Delta y = 1\,\text{nm}$ set well below $D_0 \approx 6.6\, \text{nm}$, and the cell thickness $\Delta z$ ranging from $0.5 - 4.0\,\text{nm}$. Since there is always only one cell in the $z$ direction, this enforces the assumption that $\fet m$ is independent of $z$. We used periodic boundaries in the $x$ direction, and to remove the effect of the restoring force $B_R$ we continually updated the position of the simulation window along $y$ to keep the DW centered inside the window. For each film thickness $\delta = \Delta z$, simulations were carried out in a $1\,\text{mT}$ range centered on the Walker field given by Eq.~(\ref{eq:walkerfield}). The resulting average DW velocities $V_{DW}$ for each simulation are shown in Fig.~\ref{fig:vdwvsbz} (filled symbols). The DW velocity increases with the applied field $B_a$, until Walker breakdown $B_a = B_W$ when the average velocity drops abruptly due to precession. The peaks in $V_{DW}$ (open symbols in Fig.~\ref{fig:vdwvsbz}) are therefore numerical estimates of the Walker field for each film thickness. 

\begin{figure}[tp]
  \includegraphics{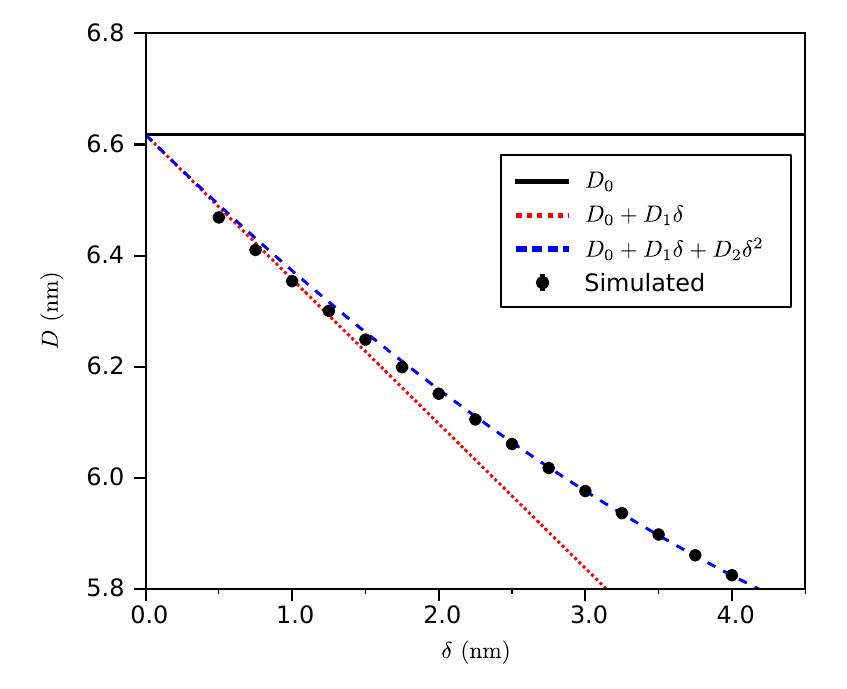}
  \caption{Numerical DW width $D$ as a function of film thickness $\delta$ (points), compared with the analytical prediction given by Eq.~(\ref{eq:D_eq}), taken to different orders in $\delta$ (lines).}
  \label{fig:dvsdelta}
\end{figure}

The DW width is measured numerically by fitting a line $ay + b$ to values of $\atanh(m_z)$ close to the DW position, which is defined as the location where $m_z$ crosses $0$, and setting $D = |a|^{-1}$. 
The resulting values are
compared with the analytical result (\ref{eq:D_eq}) in Fig.~\ref{fig:dvsdelta}. For increasing film thickness the DW width diminishes, due to the negative first-order correction $D_1$ [see Eq.~(\ref{eq:D1})]. This is mainly due to the attractive logarithmic interaction between the antiparallel domains as evident in Eq.~(\ref{eq:EdSFinal}), so that the distance $D$ between the domains is reduced as the strength of interaction increases. 

\begin{figure}
    \includegraphics{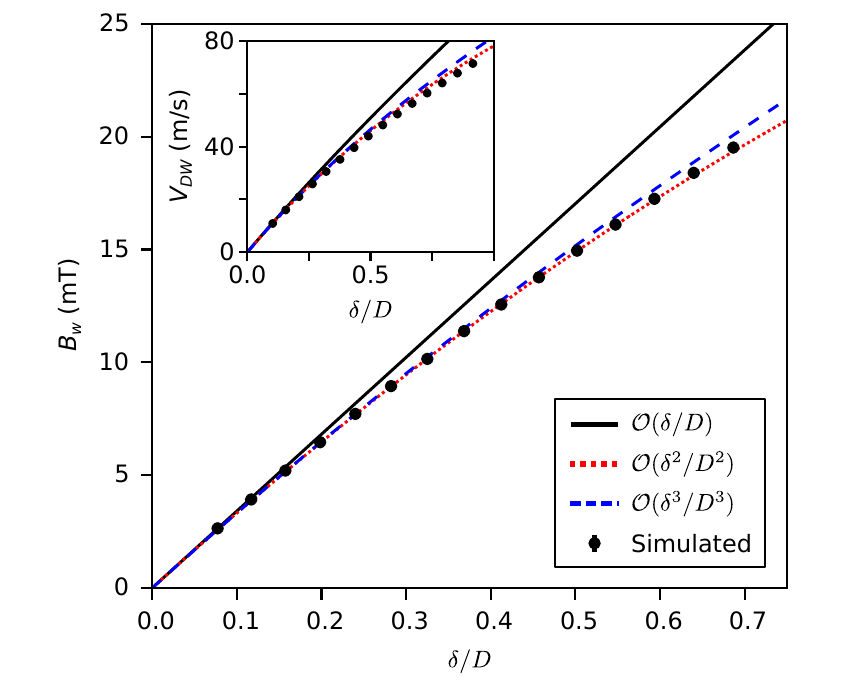}
	\caption{Numerical Walker breakdown field $B_W$ as a function of $\delta/D$ (points), compared with the analytical prediction given by Eq.~(\ref{eq:walkerfield}) including first, second, and third order terms (lines).
	Inset: Numerical Walker breakdown velocity $V_W$ (points) compared with analytical predictions given by Eq.~(\ref{eq:walkerspeed}) (lines).}
    \label{fig:Bw}
\end{figure}

The numerical values for the Walker breakdown field $B_W$ and velocity $V_W$ are shown as a function of $\delta/D$ Fig.~\ref{fig:Bw}, and compared with analytical predictions given by Eqs.~(\ref{eq:walkerfield}) and (\ref{eq:walkerspeed}) to first, second, and third order. The numerical and analytical values show good agreement with each other, highlighting the impact of higher-order corrections to the Walker field, particularly for larger values of $\delta/D$.

\begin{figure}[tp]
  \includegraphics{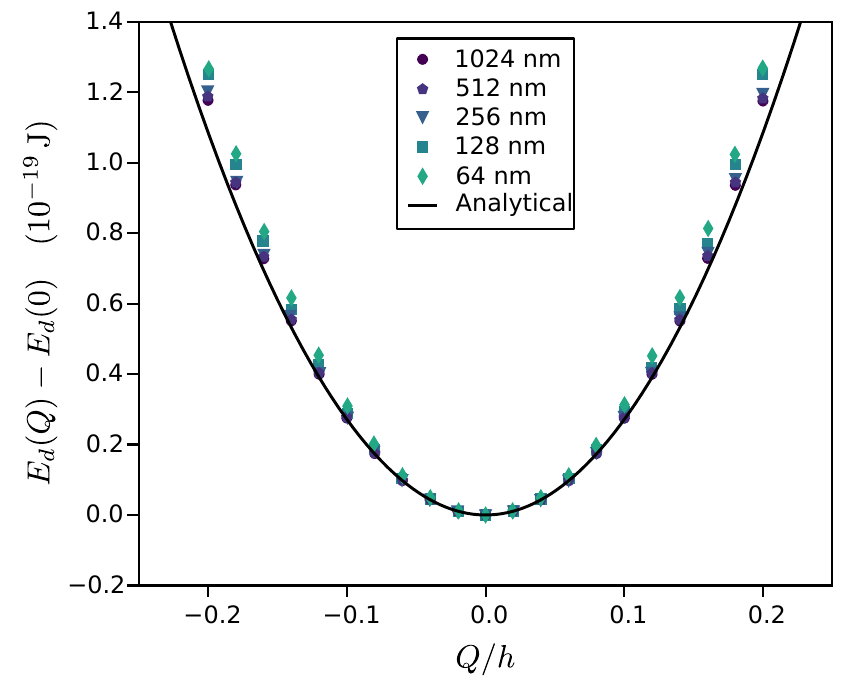}
  \caption{Numerical out-of-plane demagnetizing energy $E_d^S$ as a function of DW position $Q/h$ (points), subtracting the value at $Q = 0$ to isolate the $Q$-dependent part, compared with the analytical prediction given by
	Eq.~(\ref{eq:restoring_quadratic}) (line).}
  \label{fig:restoring}
\end{figure}

Analytical predictions for the restoring force arising from demagnetizing effects were also validated by comparison with simulations on a grid of $128 \times n_y \times 1$ cells with $64 \le n_y \le 512$, and cell size $\Delta x = \Delta y = 1\,\text{nm}$ and $\Delta z = 4\,\text{nm}$. The Bloch wall initial condition was relaxed in an applied magnetic field of varying strength, displacing the DW from the center of the sample. We then computed the demagnetizing energy of the relaxed configuration, and subtracted the $Q=0$ value to isolate the quadratic dependence given by Eq.~(\ref{eq:restoring_quadratic}). The result is given in Fig.~\ref{fig:restoring}, and shows a good agreement with the analytical prediction, up to some deviation from the quadratic behavior at large $Q^2$.

\section{Conclusion}
\label{sec:conclusion}
The magnetostatic energy of domain structures is a challenging mathematical problem in the general case. Here we have made progress on the idealized case of a uniform, infinitely long DW in a thin PMA film by deriving analytic expressions for the energy. This allows accurate predictions for important properties of the DW, such as the DW width (Sec.~\ref{sec:dwwidth}), DW dynamics (Sec.~\ref{sec:1Dmod}), and restoring force (Sec.~\ref{sec:outofplane}). Micromagnetic simulations were employed to verify our results.

The effect of the in-plane magnetization of the DW can be understood by using an effective demagnetizing constant $N_y$, whose precise value differs from the commonly used elliptic approximation. The out of plane energy, on the other hand, consists of separate contributions from the DW itself, the two domains, and the interaction between the domains, which would be difficult to disentangle from each other without the principled approach employed here. For example, the DW width is affected by the attractive interaction between the domains at either side, which the formalism of demagnetization constants fails to include. Using an explicit expression for the in-plane energy when finding the equilibrium DW width also avoids a subtle mistake due to the variation of the effective demagnetization constant $N_y$ with the DW width $D$.

For simplicity we considered the case of an infinitely extended thin film without vertical boundaries. This idealized case does not include the effect of disorder, which inevitably distort the shape of the DW in real thin films. The more realistic problem of a uniform DW in a nanostrip of width $w = \mathcal O(D)$ introduces further difficulties. Additional boundary integrals will need to be included in Eqs.~(\ref{eq:EdV}--\ref{eq:EdS}). More severely, the technique of deforming the integration contour to integrate term by term in Eqs.~(\ref{eq:EdV_deformed}, \ref{eq:EdS_deformed}) relied on simplifying the interaction kernels $g_\sigma$ and $f_\sigma$ into a simple logarithmic form [see Eqs.~(\ref{eq:Gs1}--\ref{eq:Gs3}) and (\ref{eq:Fs1})], which was obtained by taking the film width $w$ to infinity. It is unclear how to perform a similar expansion without this simplification. Some mathematical difficulties therefore stand in the way of extending these results to the case of nanostrips.

In order to consider thicker films with width $\delta \sim D$, it is necessary to allow for variations in the vertical direction $\fet e_z$ as well. In this case, the magnetization vector can deflect away from the vertical close to the surface, in order to reduce the energy penalty from the out of plane demagnetizing field. This can lead to the formation of complicated structures such as Bloch lines at the surface \cite{hubert2008magnetic,herranen2017blochline}. Our analysis is therefore restricted to $\delta \ll D$, which avoids these complications.

Another avenue for generalization is to include effects such as the Dzyaloshinskii-Moriya interaction (DMI) and spin-transfer torque. The DMI is a local energy term which can be straightforwardly included in our analysis \cite{Thiaville_2012}.
Spin-transfer torque, on the other hand, is a dynamical forcing mechanism and not an energy. It can however be included in the Lagrangian framework as a dynamical term in the Lagrangian \cite{thiaville2004domain}.

In all, our analytical computations provide a much better understanding of the effect of long-range demagnetizing fields on the properties and motion of DWs in thin films. 
The methods we have used are quite general, and can maybe be employed to understand other similar systems, such as DWs in systems with in-plane magnetic anisotropy,
and provide a solid foundation on which a principled understanding of more complicated DW behavior can be built. 

\begin{acknowledgments}
  This work has been supported by the Academy of Finland through an Academy Research Fellowship (L.L.; project no.\ 268302).
\end{acknowledgments}

\bibliographystyle{apsrev4-2}
\bibliography{ref}

\end{document}